\documentclass[twocolumn,english,showpacs,superscriptaddress]{revtex4-1}
\usepackage{geometry}
\usepackage{ragged2e}
\usepackage{xcolor}
\usepackage{babel}
\usepackage{amsmath}
\usepackage{amssymb}
\usepackage{stmaryrd}
\usepackage{graphicx}
\usepackage{wasysym}
\usepackage{bbold}
\usepackage{mathtools}
\usepackage{blkarray, bigstrut}
\usepackage{float}
\usepackage{siunitx}
\usepackage[bookmarks=false]{hyperref}
\newcommand{\ket}[1]{\vert{#1}\rangle}
\newcommand{\up}{\mid\uparrow\rangle}
\newcommand{\bra}[1]{\langle{#1}|}
\newcommand{\down}{\mid\downarrow\rangle}

\newcommand{\Ba}{$^{138}$Ba$^+$}

\newcommand{\shalfdown}{\ket{^2S_{1/2}, m_J =-\frac{1}{2}}}

\newcommand{\dhalfdown}{\ket{^2D_{5/2}, m_J =-\frac{1}{2}}}
\newcommand{\dhalfup}{\ket{^2D_{5/2}, m_J =\frac{1}{2}}}

\begin{document}
\title{High-fidelity remote entanglement of trapped atoms mediated by time-bin photons}
\date{\today}

\author{Sagnik Saha}
\email{Corresponding author: sagnik.saha@duke.edu}
\affiliation{Duke Quantum Center, Departments of Electrical and Computer Engineering and Physics, Duke University, Durham, NC 27708}
\author{Mikhail Shalaev}
\affiliation{Duke Quantum Center, Departments of Electrical and Computer Engineering and Physics, Duke University, Durham, NC 27708}
\author{Jameson O'Reilly}
\email{Present Address: Department of Physics, University of Oregon, Eugene, OR 97331}
\affiliation{Duke Quantum Center, Departments of Electrical and Computer Engineering and Physics, Duke University, Durham, NC 27708}
\author{Isabella Goetting}
\affiliation{Duke Quantum Center, Departments of Electrical and Computer Engineering and Physics, Duke University, Durham, NC 27708}
\author{George Toh}
\affiliation{Duke Quantum Center, Departments of Electrical and Computer Engineering and Physics, Duke University, Durham, NC 27708}
\author{Ashish Kalakuntla}
\affiliation{Duke Quantum Center, Departments of Electrical and Computer Engineering and Physics, Duke University, Durham, NC 27708}
\author{Yichao Yu}
\affiliation{Duke Quantum Center, Departments of Electrical and Computer Engineering and Physics, Duke University, Durham, NC 27708}
\author{Christopher Monroe}
\affiliation{Duke Quantum Center, Departments of Electrical and Computer Engineering and Physics, Duke University, Durham, NC 27708}

\begin{abstract}
Photonic interconnects between quantum processing
nodes are likely the only way to achieve large-scale quantum computers and networks. 
The bottleneck in such an architecture is the interface between well-isolated quantum memories and flying photons.
We establish high-fidelity entanglement between remotely separated trapped atomic qubit memories, mediated by photonic qubits stored in the timing of their pulses. 
Such time-bin encoding removes sensitivity to polarization errors, enables long-distance quantum communication, and is extensible to quantum memories with more than two states.
Using a measurement-based error detection process and suppressing a fundamental source of error due to atomic recoil, we achieve an entanglement fidelity of 97\% and show that fidelities beyond 99.9\% are feasible.

\end{abstract}

\maketitle

High-quality and fast photonic links between good quantum memories are crucial for large-scale modular quantum computers and networks \cite{Awschalom2021}.
Trapped atomic ions set the standard for quantum memory, with indefinite idle quantum coherence times \cite{Wang2021}, near-unit efficiency state-preparation-and-measurement (SPAM) \cite{Ransford2021}, and the highest-fidelity local quantum gates \cite{Harty2014,Ballance2016, Srinivas2021,Clark2021}.
Shuttling between multiple zones on a single trap may allow scaling to hundreds or thousands of qubits \cite{Kielpinski2002,Moses2023}, but scaling beyond this in a modular fashion will likely require optical photonic interconnects regardless of the qubit platform \cite{Awschalom2021}. 

Qubits stored in trapped atomic ion arrays are natural single photon emitters and have led the way in remote qubit entanglement protocols \cite{Moehring2007E, Hucul2015,Stephenson2020, Oreilly2024}.
In these demonstrations, photonic qubits emitted from and entangled with their parent trapped ions are interfered and detected in an entanglement-swapping procedure \cite{EntSwap1993} that leaves the two trapped ion memories entangled \cite{barrett_efficient_2005,moehring_quantum_2007, duan_colloquium_2010}.
Similar protocols have been demonstrated with neutral atoms \cite{Ritter2012, Weinfurter2022} and solid-state emitters \cite{Bernien2013, lukin_siv_entanglement_2024, ruskuc2024scalable}. 

Photonic qubits are commonly encoded in the polarization degree of freedom, allowing qubit manipulation and diagnosis with simple polarization optics. 
Remote trapped ions have been entangled through photonic polarization qubits with a fidelity of $F = 0.960(1)$ \cite{Nadlinger2022}, an entanglement rate of 250~s$^{-1}$ \cite{Oreilly2024}, and over a distance of 230 m \cite{Northup2023}. Remote neutral atoms have been similarly entangled through polarization photonic qubits with a post-selected fidelity of 0.987(22)~\cite{Ritter2012} and a rate of 0.004~s$^{-1}$.
However, polarization qubits are susceptible to uncontrolled birefringence in optical elements, windows, and optical fibers, limiting performance in these and other experiments. As system complexity grows by extending to nearby local qubit memories, additional optical elements, more connected nodes, longer distances between nodes, or higher-dimensional quantum memories, polarization qubits become even less tenable~\cite{Bersin2024, lukin_siv_entanglement_2024}. Alternative photonic degrees of freedom such as frequency \cite{Maunz2009} or time-bin~\cite{Bernien2013} are therefore preferable for a scalable architecture.

Here, we report for the first time the entanglement of two remote individual atomic qubits via time-bin photons, with a measured Bell state fidelity of ${F = 0.970(4)}$. 
Visible time-bin encoded photons emitted from remote trapped ions are collected into single-mode optical fibers, and their interference and coincident detection heralds the entanglement of the ion memories~\cite{barrett_efficient_2005}. 
We identify and suppress fidelity limits from the timing of the atomic excitations as well as the random times of photon detection.
We also flag erasure errors from the atomic qubits to purify the resulting Bell state with little overhead ~\cite{mingyu_erasure_2023}.
This demonstration shows that the fidelity limits for remote entanglement based on photons can be better than $0.999$, allowing modular scaling of quantum computers based on atomic qubits, long-distance quantum communication between quantum nodes, and the entanglement of high-dimensional quantum memories \cite{Wang2020}.

\begin{figure*}[tb]%
\centering
\includegraphics[width=0.99\textwidth]{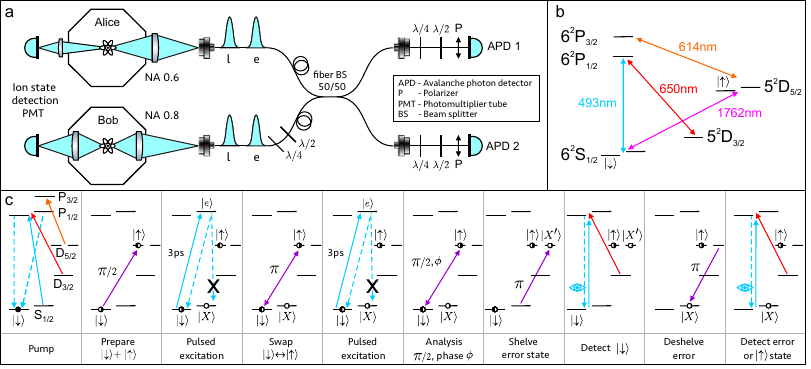}
\caption{(a) Schematic of the apparatus, including two ion trap chambers Alice (A) and Bob (B). Here e(l) denotes a photon in the early(late) time-bin. Photons collected with high numerical aperture (NA) objectives are interfered on a beamsplitter (BS) and detected with avalanche photodiodes (APD). For subsequent atomic qubit state readout, ion fluorescence is detected on photomultiplier tubes (PMT). (b) Relevant energy levels of the \Ba system (nuclear spin $I=0$). The atomic qubit is stored in the ${\down_q= \shalfdown}$ and ${\up_q= \dhalfup}$ states and coherently driven by narrowband laser radiation at 1762~nm. State preparation and measurement (SPAM) is provided through optical pumping beams at 493 nm, 650 nm, and 614 nm and Doppler laser-cooling is performed on the 493~nm and 650~nm transitions. Single photons are generated with ultrafast laser pulses at 493 nm (see Supp. Info. \ref{sec:qubit}, \ref{sec:fast}). (c) Procedure applied to each \Ba~ion at A and B for heralded remote entanglement using time-bin encoded photons.}
\label{fig:leveldiagram}
\end{figure*}

\section*{Time-Bin Entanglement Protocol}
We entangle two $^{138}\textrm{Ba}^+$ ions, each trapped in separate vacuum chambers a few meters apart, henceforth referred to as Alice (A) and Bob (B). The experimental schematic is shown in Figure~\ref{fig:leveldiagram}a.
Each chamber contains a four-rod radiofrequency Paul trap loaded with a single \Ba~ion with energy levels indicated in Fig.~\ref{fig:leveldiagram}b.
The atomic qubit $q \in \{A,B\}$ is encoded in the states ${\down_q= \ket{^2S_{1/2}, m_J =-\frac{1}{2}}}$ and ${\up_q= \ket{^2D_{5/2}, m_J =-\frac{1}{2}}}$. 

The remote entanglement protocol proceeds following Fig.~\ref{fig:leveldiagram}c. Each ion is first laser-cooled and initialized in the $\ket{\downarrow}_q$ state through optical pumping, then prepared in the superposition state $\down_q + \up_q$ 
by driving a $\pi/2$-pulse between the qubit states with 1762~nm radiation \cite{Blinov_1762_2010} (see Supp. Info. \ref{sec:qubit}). At time $t_e$, population in the $\down_q$ state is then driven with a probability $P_{\mathrm{exc}}>0.80$ to the excited state ${\ket{e}_q=\ket{^2P_{1/2}, m_J =+\frac{1}{2}}}$ using a circularly-polarized 493~nm ultrafast laser pulse (see Supp. Info. \ref{sec:fast}). 
With probability $73\%$, the spontaneous emission from $\ket{e}_q$ produces a single 493 nm photon wavepacket distributed over time given an exponential radiative lifetime of $\tau_R = 7.85$~ns \cite{Arnold2019}, and 2/3 of those decays return the population to $\down_q$ for a net branching ratio of $\beta=49\%$.
Decay to the other ground state ($\ket{^2S_{1/2}, m_J =+\frac{1}{2}}$) via a $\pi$-polarized photon is rejected with a polarization filter. Decay to the $^2D_{3/2}$ state ($27\%$ branching ratio) by emission of a 650~nm photon is rejected through spatial and spectral filtering.

The ideal unnormalized state of each ion $q$ and its collected photon mode is now 
\begin{equation}
    \sqrt{1-p_q}\down_q\ket{0_e}_q+\sqrt{p_q}e^{i\phi_{qe}}\down_q\ket{1_e}_q
    +\up_q\ket{0_e}_q
\end{equation} 
where $  \ket{1_e}$ ($\ket{0_e}$) denotes the presence (absence) of a photon in the first (early) time-bin and $p_q$ is the probability a single photon has been collected. The phase $\phi_{qe} = \mathbf{\Delta k}\cdot\mathbf{r}_q(t_e) + \phi^*_{qe}$ includes the position $\mathbf{r}_q(t_e)$ of ion $q$ at time $t_e$ of the early time-bin, and ${\mathbf{\Delta k}=\mathbf{k}'-\mathbf{k}}$ is the difference between the excitation pulse wavevector $\mathbf{k}'$ and that of the emitted photon $\mathbf{k}$ (both of magnitude $k$). The small random phase $\phi^*_{qe}$ accounts for the narrow distribution of emission times and is discussed below.

To generate the second (late) time-bin photon, the populations $\down_q$ and $\up_q$ are swapped with a 1762~nm $\pi$-pulse (see Supp. Info. \ref{sec:qubit}), then at time $t_l$ the $\down_q$ state is again excited to the $\ket{e}_q$ state.
With probability $p_q$, there is now a single-photon time-bin qubit entangled with its parent ion qubit, ideally in the state
\begin{equation}e^{i\phi_{qe}}\up_q\ket{1_e0_l}_q
     +e^{i\phi_{ql}}
     \down_q\ket{0_e1_l}_q,
\end{equation} 
where $\ket{n_e n_l}_q$ denotes a state of $n_e$ ($n_l$) photons in the early (late) time-bin from ion $q$.

The photons are then directed to a non-polarizing 50:50 fiber beamsplitter (BS), which erases their ``which-path" information through Hong-Ou-Mandel interference \cite{HOM1987}. 
Subsequent detection of early and late photons ideally projects the ions into a Bell state \cite{Moehring2007E}
\begin{equation}
    \Psi^{\pm}=\down_A\up_B \pm e^{i \phi}\up_A\down_B \label{eqn:state}
\end{equation}
where the phase is ${\phi = (\phi_{Ae} - \phi_{Be}) - (\phi_{Al} - \phi_{Bl})}$.   
The $\Psi^{+}$ $(\Psi^{-})$ state is heralded by early and late detections on the same (opposite) BS output channels. 

\begin{figure}[tb]%
\centering
\begin{tabular}{|c|c|c|c|c|c|} \hline 
      Bell state&$P_{\uparrow_A\downarrow_B}+P_{\downarrow_A\uparrow_B}$& $C$ & $F$  &$\delta t$ & $Y$\\ \hline \hline
    $\Psi^{+}$& 0.990(4)& 0.927(6)& 0.959(4) & 50~ns & 0.998\\ \hline 
    $\Psi^{-}$& 0.996(3)& 0.931(6)& 0.963(3) & 50~ns & 0.998\\ \hline \hline
    $\Psi^{+}$& 0.990(4)& 0.948(6)& 0.968(4) & 10~ns & 0.714\\ \hline
    $\Psi^{-}$& 0.996(3)& 0.949(6)& 0.972(3) & 10~ns & 0.714\\ \hline
\end{tabular}
\includegraphics[width=0.46\textwidth]{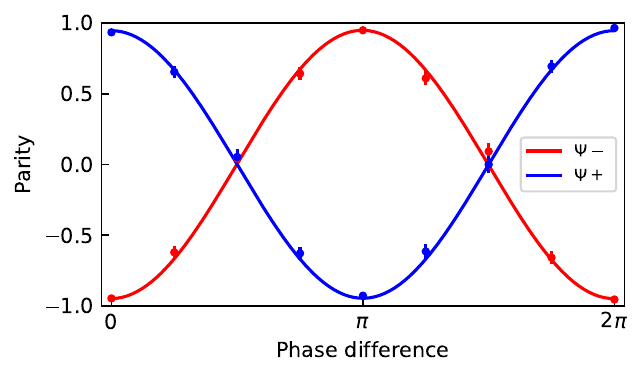}
\caption{The table shows measured populations, parity fringe contrast, and fidelities for the Bell states $\Psi^{\pm}$ for two cases of detection window $\pm\delta t$ and the corresponding yield $Y$. The plot shows the measured state parity fringe for $\Psi^-$ (red) and $\Psi^+$ (blue) following $\pi/2$ rotations of each qubit while scanning their relative phase ($\delta t = 10$ ns).}
\label{fig:parflops}
\end{figure}

\section*{Entanglement Rate and Fidelity}
The maximum success probability of ion-ion entanglement is $P_E=\frac{1}{2}p_A p_B = 2.3 \times 10^{-5}$. The factor of 1/2 accounts for the rejection of events where both photons are collected in the same time-bin (and on the same detector). 
We select coincident events occurring within a detection time window $\pm \delta t$ of the mean photon arrival time of each time-bin, resulting in a small reduction in $P_E$ by the yield factor $Y=1-e^{-\delta t/\tau_R}$, given the atomic lifetime $\tau_R$.
With a detection time window of $\delta t = 10$ ns, we observe a mean entanglement rate of ${0.35\,s^{-1}}$. A full description of photon collection parameters is provided in Supp. Info.~\ref{sec:rate calc}.

We characterize the entangled state by measuring qubit correlations in different bases. The fidelity with respect to the nominal Bell state in Eq. \ref{eqn:state} is given by $F=(P_{\mathrm{odd}} + C)/2$, where $P_{\mathrm{odd}}$ is the population of the odd parity states $\down_A\up_B$ or $\up_A\down_B$ and $C$ is the contrast of parity oscillations of the two qubit states as the relative phase of analysis $\pi/2$ rotations on each qubit is scanned \cite{Sackett2000}. Fig. \ref{fig:parflops} shows measurements for both states $\Psi^{\pm}$. 
We detect the ion qubit states by shining 493~nm and 650~nm light on the ions, which causes the ions in the $\ket{\downarrow}_q$ state to fluoresce, while ions in the $\ket{\uparrow}_q$ state remain dark. This provides deterministic qubit state detection with a fidelity exceeding 99.5$\%$ (see Supp. Info. \ref{sec:qubit}). 
As shown in Fig.~\ref{fig:parflops}, the measured fidelities of the entangled states (uncorrected for SPAM) are $F = 0.968(4)$ for the state $\Psi^+$  and $F=0.972(3)$ for the state $\Psi^-$.

We attribute most of the observed fidelity imperfection to slow drifts in the intensity of the 1762~nm qubit laser.
We observe $\sim1\%$ fluctuations over the few-hour time period of an experimental run, while slow drifts are eliminated between runs through periodic calibration.
These fluctuations degrade SPAM and are expected to contribute to a fidelity error of $1\%$.
Other sources of error include leakage via the $\pi$ decay channel and residual entanglement with motion (both discussed separately below); micromotion Doppler shifts; and optical system imperfections. The overall error budget is discussed in Supp. Info. \ref{sec:error}. 

\section*{Correction of Erasure Errors}

During the photon emission process, there is a $\sim24\%$ probability that each ion decays to the wrong ground state $\ket{X}= \ket{^2S_{1/2}, m_J =+\frac{1}{2}}$. Although the corresponding $\pi$-polarized photons are blocked by a polarizer with $>98\%$ efficiency, polarization mixing from imperfect alignment \cite{Kim2011} or drifts of the fiber birefringence make it difficult to passively eliminate these false positives. However, we can flag this qubit erasure error by shelving the state $\ket{X}$ to $\ket{X'}=\ket{^2D_{5/2}, m_J =+\frac{1}{2}}$ before state detection. After state detection, we de-shelve $\ket{X'}$ back to $\ket{X}$ and perform another round of state detection to check for the error (see Fig. \ref{fig:leveldiagram}c). 
This allows for the suppression of erasure errors to below 0.1\% \cite{Wu2022} with very little loss in success rate.
This erasure-veto technique will play an increasingly important role in suppressing errors when single mode fibers susceptible to polarization drifts are used for long-distance quantum communication~\cite{Bersin2024}.

\begin{table}
\caption{Measured harmonic motional frequencies $\omega_{qi}$ for the two atomic ions $q=A,B$ along direction $i$ and their commensurability with the photonic excitation rate $\tau^{-1}=165.35$ kHz  ($\tau=6048$~ns). The six mode frequencies are set to be nearly integer multiples of the excitation rate, suppressing errors from residual entanglement with motion. Also shown are the Lamb-Dicke parameters $\eta_{qi}$ and $\zeta_{qi}$ with respect to the excitation/emission wavevector difference and the wavevector of emission, respectively.}
\begin{tabular}{ |c|l|r|r|r|r|r| }
 \hline
  $q$ (ion) & $i$ (mode) & $\frac{\omega_{qi}}{2\pi}$ (kHz) & $\frac{\omega_{qi}\tau}{2\pi}$ & $\eta_{qi}$\phantom{0} & $\zeta_{qi}$\phantom{00} \\
  \hline
    \hline
 A & Axial    &  991.5& 5.996& 0.055 & 0 \phantom{00}\\ \hline
 A & Radial 1 & 1157.5& 7.000& 0.086 & 0.051\phantom{7}\\ \hline
 A & Radial 2 & 1488.0& 8.999& 0.013 & 0.045\phantom{7}\\ \hline
   \hline
 B & Axial    & 330.3 & 1.997& 0.095 & 0 \phantom{00}\\ \hline
 B & Radial 1 & 826.7 & 4.999& 0.066& 0.0067\\ \hline
 B & Radial 2 & 992.0 & 5.999& 0.073 & 0.077\phantom{7}\\ \hline
\end{tabular}
\label{table:modes}
\end{table}

\section*{Timing and Atomic Recoil Errors}

The fidelity of photonically-heralded atom-atom entanglement is sensitive to time differences in photon detection, owing to a betrayal of ``which-path" information from differences in atomic recoil.
After tracing over the motion of both ions, the average parity fringe contrast degrades to (see Supp. Info. \ref{sec:motion} and Refs. \cite{Johnson2015},\cite{Yichao2024})
\medmuskip=3mu
\thinmuskip=4mu
\thickmuskip=5mu
\begin{align}
C &= \prod_{qi} e^{-(2\bar{n}_{qi}+1)\left[\eta_{qi}^2(1-\cos \omega_{qi}\tau) +  \zeta_{qi}^2 W \omega_{qi}^2\tau_R^2\right]}
\label{eqn:FidLim}
\end{align}
Here, $\omega_{qi}$ and $\bar{n}_{qi}$ are the frequency and thermal phonon occupation number for  mode $i$ of ion $q$. 

\begin{figure}[t]%
\centering
\includegraphics[width=0.47\textwidth]{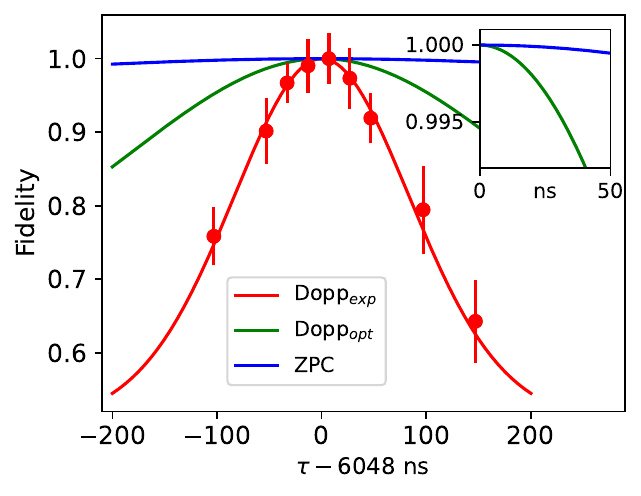}
\caption{Entanglement fidelity degradation as time-bin period $\tau$ is tuned away from being commensurate with trapped ion motion (in all three directions of motion for both ions). 
As predicted in Eq. \ref{eqn:FidLim}, the loss of fidelity due to asynchronous timing depends on the thermal occupation number of each of the six modes of motion, with the value  $\tau=6048$~ns nearly eliminating this error (see Table \ref{table:modes}).
Points are measured fidelity values, re-scaled to a maximum of 1, and the curves are theoretical fidelity limits from the first term of Eq. \ref{eqn:FidLim} at various levels of cooling. The red line corresponds to Doppler cooling in the experiment (Dopp$_{\mathrm{exp}}$) given the geometry of the cooling beams with respect to the principal axes in the experiment (see Supp. Fig. \ref{fig:angles}). The green line is for Doppler cooling with an optimal beam geometry (Dopp$_{\mathrm{opt}}$) and the blue line is for zero-point cooling.}
\label{fig:secularfrequencies}
\end{figure}

The first decoherence term in Eq. \ref{eqn:FidLim} stems from an entanglement between the qubits and the atomic motion from the separated time of excitation $\tau=t_l-t_e$ (seen in the phase of Eq. \ref{eqn:state}) and is specific to time-bin encoding schemes. The Lamb-Dicke recoil parameter is $\eta_{qi}=\Delta k_{i}\sqrt{\hbar/2m \omega_{qi}}$ for ion $q$ with mass $m$ with respect to the wavevector difference between excitation and emission along $i$. We see that when  $\omega_{qi}\tau$ is an integer multiple of $2\pi$ for all modes, each ion is excited from the same position in each time bin, and this source of decoherence vanishes. 
In the experiment, we ensure this condition by tuning the mode frequencies to be commensurate and setting the excitation rate $1/\tau$ to be their greatest common divisor ($\tau= 6048$~ns), as summarized in Table \ref{table:modes}.
We characterize this effect by scanning the difference in excitation times $\tau$ about the nominal value, as shown in Fig.~\ref{fig:secularfrequencies}. We estimate that the residual fidelity error from the drift in mode frequencies is less than $0.1\%$.

\begin{figure}[h]
\centering
\includegraphics[width=0.48\textwidth]{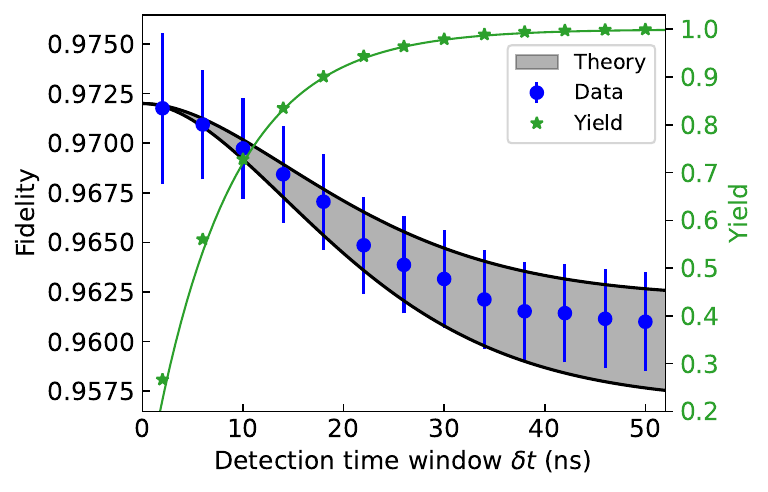}
\caption{Observed fidelity as a function of the detection time window $\delta t$ between the excitation and emission of photons from each ion. The shaded region indicates the expected fidelity including motional recoil from Eq. \ref{eqn:FidLim}, assuming Doppler cooling for all modes given the beam geometry (see Supp. Info. Fig. \ref{fig:angles} and Table \ref{table:modes_supplementary}). This band was normalized to match the experimental value at $\delta t =0$ with a width reflecting the uncertainty in the angles of the ion trap principal axes.
The green line (right axis) shows the measured (points) and theoretical (line) yield for each value of $\delta t$, with no free fit parameters.} 
\label{fig:deltat-fidelity}
\end{figure}

The second decoherence term in Eq. \ref{eqn:FidLim} is more fundamental and stems from fluctuations in the random detection times of the photons in each time bin through the random phase $\phi^*_{qe}$ given by the finite lifetime of the emitting atoms. This generates residual entanglement between the qubits and their motion as above. But in this case, the Lamb-Dicke recoil parameter ${\zeta_{qi}=k_{i}\sqrt{\hbar/2 m \omega_{qi}}}$ is with respect to the emission wavevector only and not $\Delta k_{i}$.
This decoherence can be controlled by narrowing the detection window $\delta t$ characterized by the scaled variance $0 < W < 1$ (see Supp. Info. \ref{sec:motion}), but this also degrades the yield $Y$ and hence the rate of entanglement. 
Figure~\ref{fig:deltat-fidelity} shows the observed fidelity and yield as we vary $\delta t$ from $2$ ns ($W\approx0.01$, $Y=0.22$) to $50$~ns ($W\approx0.95$, $Y=0.998$). The measurements are consistent with the model of Eq. \ref{eqn:FidLim}, assuming thermal states of motion near the Doppler laser-cooling limit for all modes.
We observe a $\sim1\%$ improvement in the fidelity by decreasing the window from 50 ns to 10 ns ($Y=0.71$), with a residual fidelity error of $\sim0.2\%$.
This decoherence from random photon arrival times is universal to all photonic encoding schemes for recoiling emitters but has not been previously observed.

We note that for fixed emitters such as color-centers in solid-state hosts \cite{Bernien2013}, the emitter mass becomes so large that $\eta_{qi},\zeta_{qi} \rightarrow 0$ and the above recoil-induced decoherence is negligible. For very weakly-bound emitters such as neutral atoms \cite{Ritter2012, Covey2023}, where $\omega_{qi}\tau, \zeta_{qi} \ll 1$, these effects can be prominent, depending on the level of cooling (see Supp. Info. \ref{sec:motion}).

\section*{Summary}
Remote entanglement of trapped ions using time-bin encoded photons is the preferred method for high fidelity or long-distance photonic interconnects between quantum memory nodes. 
By stabilizing the 1762~nm laser power and using hyperfine clock qubits with indefinitely long idle coherence times, stabilized mode frequencies on the order of MHz, low $\bar n < 1$ via sub-Doppler cooling, and improved SPAM, it should be possible to reach remote entanglement fidelities in excess of 0.999.
Furthermore, by increasing and stabilizing the intensity of the 1762~nm laser to achieve stable Rabi frequencies exceeding a MHz should enable the operation of this protocol at rates approaching $10^3$ per second.
Such high-fidelity and high-rate entanglement will be critical for scaling of a photonically networked quantum computer, as well as quantum repeaters and other long-distance quantum communication protocols \cite{Ramette2023}. 
Moreover, by extending the above protocol to any number of time-bins, this type of photonic interconnect can easily interface with high-dimensional qubit registers~\cite{Wang2021, Low2023}
to generate particular entangled qudit states for applications in networking \cite{Islam2017} and quantum computation \cite{Sun2024}.

This work is supported by the DOE Quantum Systems Accelerator (DE-FOA-0002253) and the NSF STAQ Program (PHY-1818914). J.O. is supported by the NSF Graduate Research Fellowship (DGE 2139754) and A.K. by the  AFOSR National Defense Science and Engineering Graduate (NDSEG) Fellowship.

\bibliographystyle{apsrev4-1}
\bibliography{Refs, time-bin-sagnik-1}

\clearpage
\newpage
\section*{Supplementary Materials}

\subsection{Optical qubit SPAM and coherent rotations}    
\label{sec:qubit}
The atomic qubit spans ${\down = \shalfdown}$ and ${\up = \dhalfdown}$ with an optical interval at a wavelength of 1762~nm. The linear Zeeman splitting between the two levels is $0.56$~MHz/G \cite{Kurz2010}.
A static magnetic field of approximately 4~G is applied to each chamber and balanced to under 1~mG to bring the optical qubit frequencies within 200 Hz of each other.

Coherent individual qubit rotations are performed using a 1762 nm Tm-doped, distributed-feedback fiber laser from NKT Photonics. 
The laser light is directed through a fiber amplifier for a total output of 450 mW. 
By locking to a high-finesse cavity, the laser frequency is stabilized to a linewidth of less than 200~Hz.
The free-space 1762~nm light is split into two paths and coupled into individual fiber AOMs for Alice and Bob. 
The rf waveforms that drive the two AOMs are generated by a multi-channel arbitrary waveform generator to ensure a stable relative phase between the pulses. 
In each system, we direct about 20~mW of laser power focused down to a 20~$\mu$m waist and
drive a $\pi$ transition from $\down=\shalfdown$~ to $\up = \dhalfdown$ in $\sim4.2$~$\mu$s (Rabi frequency of $\sim 120$~kHz).

\begin{figure}[h]%
\centering
\includegraphics[width=0.4\textwidth]{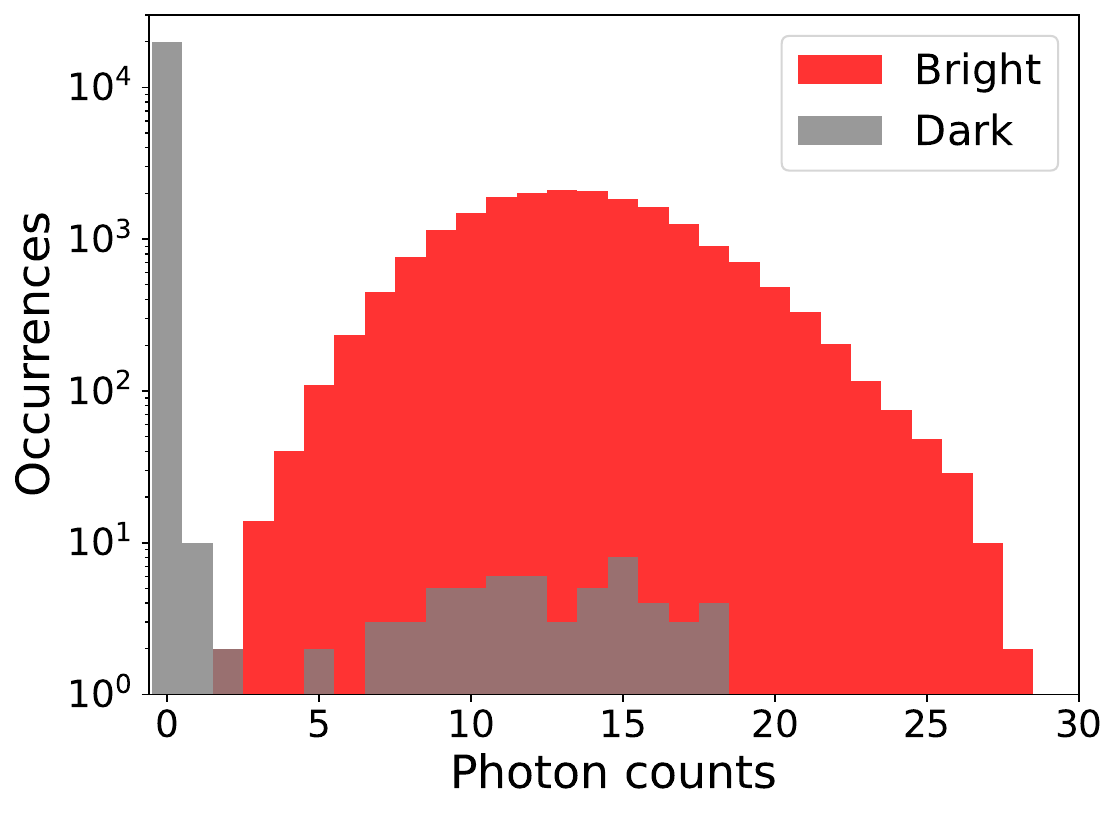}
\caption{Histogram of ion-fluorescence counts in Alice for qubit state prepared in $\down$ (dark) and $\up$ (bright). The fluorescence integration time is 1~ms per measurement, averaged over 20,000 repetitions. Setting a threshold of 2.5 counts allows the two qubit states to be distinguished in a single shot with measurement fidelity $>99.5\%$. }
\label{fig:1762histogram}
\end{figure}

State preparation involves optical pumping into $\down$ with $\sigma^-$-polarized 493~nm light, and 650~nm and 614~nm re-pump beams to clear the D manifolds (see Fig.~\ref{fig:leveldiagram}b). 
Measurement is performed by ion-fluorescence-based detection. 
Since $^2D_{5/2}$ has a lifetime of $\tau\approx30s$, the $\up$ state is dark when the ion is exposed to 493~nm and 650~nm light, while the $\down$ state fluoresces, emitting photons that are counted by a photomultiplier tube (PMT) detector. 
We perform fluorescence detection for a period of 1~ms, with histograms of fluorescence counts shown in Fig.~\ref{fig:1762histogram} for Alice. 
To differentiate between bright and dark states, we use a threshold of 2.5 counts and 10.5 counts for Alice and Bob, respectively. 
Employing this method, we achieve a state preparation and measurement~(SPAM) fidelity exceeding $99.5\%$ for both Alice and Bob with the residual SPAM error dominated by fluctuations in the 1762 nm operations. None of the data presented here is corrected for SPAM errors.

\subsection{Measurement of qubit coherence time ($T_2^*$)}
\label{sec:T2}
\begin{figure}[h]
\centering
\includegraphics[width=0.4\textwidth]{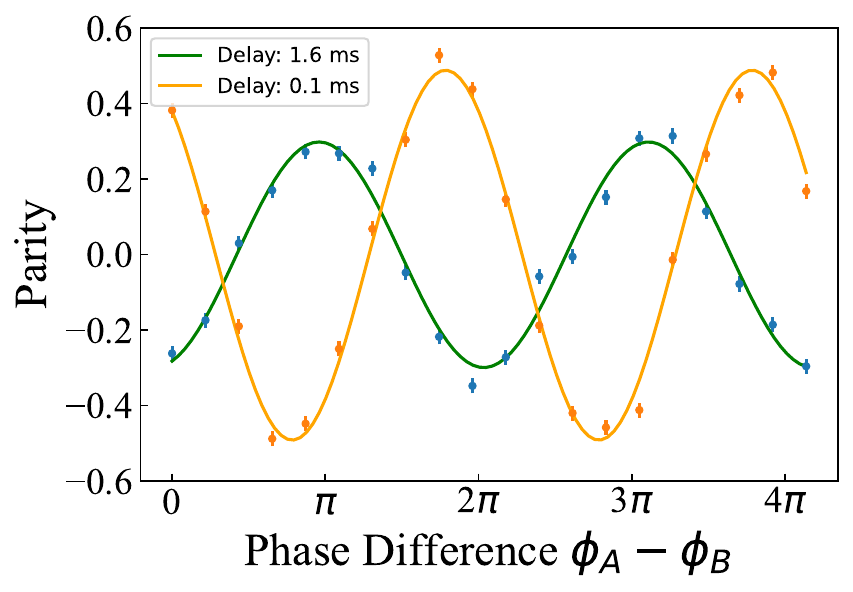}
\includegraphics[width=0.4\textwidth]{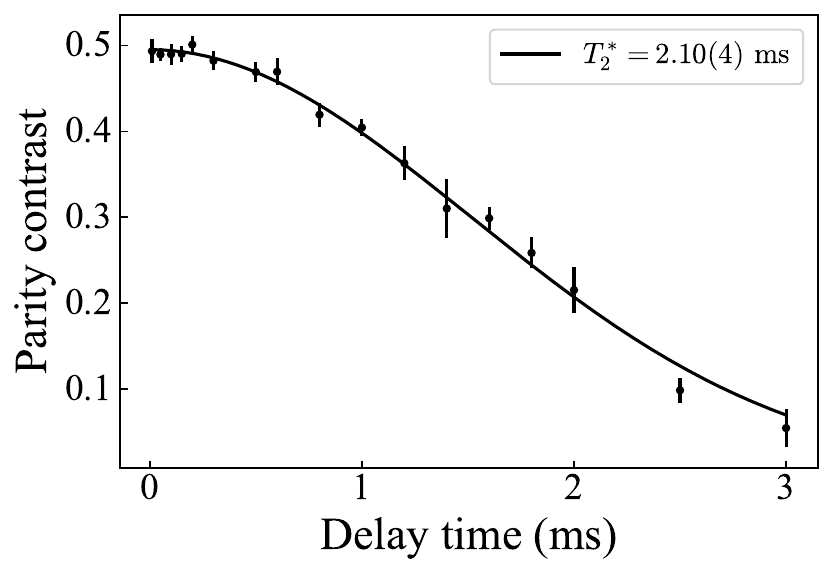}
\caption{Measurement of differential qubit coherence time between the qubits in Alice and Bob. a Measured oscillations of the two qubit parity with different relative phase for delay times of 0.5~ms and 1.6~ms. (b) Parity contrast as we vary the delay time between two $\pi/2$ pulses. Because the two ions are not entangled, a maximum parity contrast of 0.5 is expected. The fit to a Gaussian envelope gives a $1/e$ coherence time of $T_2^* = 2.10(4)$ ms.}
\label{fig:acramsey}
\end{figure}

The coherence of the desired entangled state of Eq. \ref{eqn:state} is insensitive to common mode qubit decoherence. In order to determine the effect of differential qubit decoherence on the observed state fidelity (presumably from differential magnetic field noise), we perform a Ramsey experiment on qubit A as observed from the frame of qubit B~\cite{chwalla2007, Olmschenk2007}. After preparing both ions in the $\down$ state, we use the 1762~nm laser to apply a $\pi/2$ pulse to both qubits, wait for some delay time $\Delta t$, and then apply another $\pi/2$ pulse with a scrambled phase.
By varying the relative phase difference of the second pulse between the two ions, we measure and fit the parity of the two qubit states to extract the contrast as shown in Fig.~\ref{fig:acramsey}a. 
Since the two qubits are not initially entangled, the maximum parity expected without decoherence is 0.5.
After repeating the measurement for different $\Delta t$, we fit the parity amplitudes to measure the differential decoherence time $T_2^*$. 
From the data shown in Fig.~\ref{fig:acramsey}b, we fit to $\exp[-(t/T_2^*)^2]$ and obtain $T_2^* = 2.10(4)$ ms. This is consistent with an rms differential magnetic field noise of $\sim 1$ mG over the measurement bandwidth.

\subsection{Entanglement rate calculation} 
\label{sec:rate calc}
The maximum success probability of ion-ion entanglement is $P_E=\frac{1}{2}p_A p_B = 2.3 \times 10^{-5}$.
The individual collection and detection probabilities of each node are $p_q=p_{\mathrm{exc}} \beta \epsilon_F T \epsilon_D (d\Omega_q/4\pi)$. 
Here $p_{exc}=0.8$ is the probability of excitation, $\beta=0.49$ is the effective branching ratio into the correct subspace,
$\epsilon_F \approx 19\%$ is the fiber coupling efficiency (including polarization rejection by the fiber \cite{Kim2011}), $T \approx 90 \%$ is the transmission through optical elements, $\epsilon_D=0.71$ is the detector efficiency of the avalanche photodiodes (APD),
and $d\Omega_q$ is the solid angle of light collection from chamber $q$. 
The objective lenses have numerical apertures (NA) of 0.6 in Alice and 0.8 in Bob \cite{Carter2024}), so $d\Omega_A/4\pi= 10\%$ and $d\Omega_B/4\pi = 20\%$. For a detection time window of $\delta t =10$ ns (yield $Y=0.71$), the mean entanglement rate is ${P_E Y R f =0.35\,s^{-1}}$ at a repetition rate of $R\approx70$ kHz and a duty cycle of $f=30\%$ due to laser-cooling interruptions.

\subsection{Photon generation with ultrafast 493 nm laser}
\label{sec:fast}
Single photon generation relies on excitation pulses of duration $t_p \ll \tau_R$, where $\tau_R = 7.855$ ns is the radiative lifetime of the excited $\ket{e}$ state. 
To generate fast, solitary pulses at 493 nm, we use a mode-locked Ti:Sapphire laser (Coherent Mira 900P) at 986~nm, producing $t_p\approx 3$~ps pulses at a repetition rate of $f_{\mathrm{rep}}=76.226$~MHz~\cite{Oreilly2024}. The pulses are sent through an electro-optic pulse picker that selectively transmits a single pulse when triggered. 
These pulses are then frequency-doubled with a MgO-doped, periodically-poled lithium niobate crystal to 493~nm, quadratically increasing the extinction ratio of neighboring pulses to a level below $10^{-4}$. After passing through an AOM that further extinguishes subsequent pulses, the single pulse is split in two paths and fiber-coupled into polarization-maintaining optical fibers directed to Alice and Bob.

The probability of two photons being emitted from either atom after a single pulse is estimated to be
$P_{\mathrm{exc}}^2\beta^2(t_p/8\tau_R) < 10^{-5}$. Here $P_{\mathrm{exc}} \sim 0.8$ is the excitation probability over the duration of the pulse and $\beta=0.49$ is the successful branching ratio back to the $\down$ state.

We connect the clock output of the Ti:Sapphire laser to the control system to synchronize the start of the experimental procedure with the laser repetition rate. 
This synchronization leads to a more precise time-stamp of the arrival photons, removing a potential unsynchronized $1/f_{\mathrm{rep}}\approx13$~ns timing jitter.

\subsection{Fidelity limits from atomic recoil over time} \label{sec:motion}
Multiple excitation times in the time-bin protocol or even the distribution of emission times within a single time-bin can lead to entanglement between the photon qubit and motion of the each atom from atomic recoil.
We model the resulting decoherence on the atomic qubits by calculating the reduced density matrix of each ion and its emitted photon and tracing over the motion \cite{Yichao2024}.
We first consider the temporal separation of the time bins and neglect the random distribution of emission times by taking $\phi^*_q=0$.

The initial state of ion $q \in (A,B)$, including its time-bin photon emission modes and motion is
\medmuskip=1mu
\thinmuskip=2mu
\thickmuskip=3mu
\begin{equation}
\rho_q =\frac{(\up_q+\down_q)(\bra{\uparrow}_q+\bra{\downarrow}_q)}{2}\otimes \ket{0_e0_l}_q \bra{0_e0_l}_q\otimes M_{q},
\end{equation}
where $|N_e N_l\rangle_q$ denotes $N_e$ ($N_l$) photons emitted in the early (late) time bin. The initial motional density matrix $M_q$ is expressed as a thermal state in the basis of coherent states \cite{Glauber1963}:
\begin{equation}
M_{q}=\prod_{i}\frac{1}{\pi\bar{n}_{qi}}\int\mathrm{d}^2\alpha_i|\alpha_i\rangle\langle\alpha_i|e^{-|\alpha_i|^2/\bar n_{qi}},
\end{equation}
where $\bar n_i$ is the average thermal motional quantum number in direction $i$.

The final state after the early ($e$) excitation, qubit swap, and free evolution in the ion trap for time $\tau$ between photon emissions is
\begin{equation}
\rho_q' = (L_{q}e^{-iH_q\tau} X_q E_{q})\rho_q (L_{q}e^{-iH_q\tau} X_q E_{q})^\dagger.
\end{equation}
Here, the Hamiltonian $H_q=\sum_{i}\omega_{qi}(n_{qi}+\frac12)$ describes the free evolution of the atomic motion in the trap with harmonic frequencies $\omega_{qi}$ and phonon occupation numbers $n_{qi}$ in all three dimensions. The Pauli spin-flip operator $X_q$ describes the qubit swap between emission attempts. 
The production of a photon from ion $q$ in the early or late time-bin is given by the evolution operators
\begin{align}
E_{q} &= \down_q\bra{\downarrow}_q 
\left(\sqrt{p_q}e^{i\mathbf{\Delta k}\cdot \mathbf{r}_q}a^\dagger_{q}+ \sqrt{1-p_q} \right) + \up_q\bra{\uparrow}_q \label{eqn:emissionE} \\
L_{q} &= \down_q\bra{\downarrow}_q 
\left(\sqrt{p_q}e^{i\mathbf{\Delta k}\cdot \mathbf{r}_q}b^\dagger_{q}+ \sqrt{1-p_q} \right) + \up_q\bra{\uparrow}_q, \label{eqn:emissionL}
\end{align}
where $p_q$ is the success probability of photon collection, 
$\mathbf{\Delta k}$ is the difference in the wavevector between the excitation and emitted photons, $\mathbf{r}_q$ is the (time-independent) atomic position operator of ion $q$, and $a^\dagger_{q}$ ($b^\dagger_{q}$) is the creation operator for ion $q$ producing a photon in the early (late) time-bin mode. 
The phase factor in Eqs. \ref{eqn:emissionE}-\ref{eqn:emissionL} is recognized as a momentum kick $e^{i\mathbf{\Delta k}\cdot \mathbf{r}_q}=\prod_{i}\mathcal{D}_i(i\eta_{qi})$,
where $\mathcal{D}_i$ is the coherent displacement operator in the $i$th dimension of phase space \cite{Glauber1963} and ${\eta_{qi}=\Delta k_{i}\sqrt{\hbar/2 m \omega_{qi}}}$ is the Lamb-Dicke parameter of ion $q$ associated with $\Delta k_{i}$. 

By tracing over the motion we find that with probability $p_q$, 
the reduced density matrix for qubit $q$ and its photonic channel becomes the mixed state
\begin{align}
\mathrm{tr}_{M_q}({\rho'_q}) =
\renewcommand\arraystretch{2}
\left[
\begin{array}{cc}
  1/2 &  e^{-i\phi_{q0}}C'_q/2 \\
    e^{i\phi_{q0}}C'_q/2 & 1/2 
\end{array}
\right]
\end{align}
written in the basis of the two states $\down_q\ket{0_e1_l}_q$ and $\up_q\ket{1_e0_l}_q$, where the coherence amplitude is
\begin{align}
C' &= \prod_{q} C'_{q}= \prod_{qi}e^{-\eta_{qi}^2(2\bar{n}_{qi}+1)(1-\cos\omega_{qi}\tau)} \label{eqn:contrastsupp}
\end{align}
and its zero-point phase offset is $\phi_{q0} = \eta_{qi}^2\sin \omega_{qi} \tau$, which is very small for $\eta_{qi} \ll 1$.

In addition to the reduction in coherence from the time-bin separation $\tau$ of excitation pulses, there is another reduction from the random times of photon detection. This stems from the finite lifetime $\tau_R$ of each atomic emitter resulting in the random phase $\phi^*_{qe}$ that appears in Eq. \ref{eqn:state}.  
Similar to the above treatment, we find that the coherence amplitude is reduced further by the factor
\begin{align}
&C''_q = \prod_{i}e^{-\zeta_{qi}^2(2\bar{n}_{qi}+1)[1-\cos\omega_{qi}(\tau^*-\tau)]} , \label{eqn:randomsupp}
\end{align}
with an additional negligible phase offset \cite{Yichao2024}. Here, $\tau^*$ is the measured difference in detection time of the two photons in the early and late time bins for each event. (This assumes balanced optical path lengths after the BS, but any imbalance can be factored into the data analysis with no degradation.) This is similar to the form of Eq. \ref{eqn:contrastsupp}, except the relevant Lamb-Dicke parameter ${\zeta_{qi}=k_{i}\sqrt{\hbar/2 m \omega_{qi}}}$ is associated with only the recoil from emission. The random variable $\tau^*$ follows a double-sided exponential (Laplace) distribution with mean $0$ and variance $2\tau_R^2$. 
We can reduce the impact of this variance by symmetrically truncating this distribution by post-selecting events with photon detection times within $\pm\delta t$ of the nominal value, or $|\tau^*-\tau| < \delta t$.
This results in $\tau^*-\tau$ following a truncated Laplace distribution with mean $0$ and variance $2\tau_R^2 W$, where the variance parameter
\begin{align}
W=\frac{1-(1+w+w^2/2)e^{-w}}{1-e^{-w}}  
\end{align}
smoothly increases from 0 to 1 as the relative window size $w \equiv \delta t/\tau_R$ increases from 0 to $\infty$. The yield of accepted events is $Y=1-e^{-w}$. 
For $\omega_{qi}\tau_R \ll 1$, the above average results in
\begin{align}
C''  &\approx \prod_{qi} e^{-\zeta_{qi}^2(2\bar{n}_{qi}+1) W \omega_{qi}^2\tau_R^2}.
\end{align}
Furthermore, if also $\zeta_{qi}^2(2\bar{n}_{qi}+1)\omega_{qi}\tau_R \ll 1$, then
\begin{align}
C''&\approx 1- \sum_{qi}\zeta_{qi}^2(2\bar{n}_{qi}+1) W \omega_{qi}^2\tau_R^2.
\label{eqn:FidLimS}
\end{align}
The net entanglement contrast is $C=C' C''$, as written in Eq. \ref{eqn:FidLim} in the main body.

\subsection{Beam geometry and Lamb-Dicke Parameters for optimal cooling}
\label{sec:geometry}
\begin{figure}[b]
\centering
\includegraphics[width=0.45\textwidth]{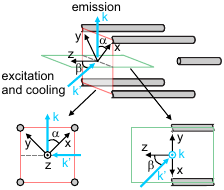}
\caption{The orientation of trap principal axes $x$, $y$ and $z$ with respect to pulsed excitation, Doppler cooling and single photon emission directions (see Table \ref{table:modes_supplementary}). The excitation and Doppler cooling wavevector $k'$ is perpendicular to the emission wavevector $k$. }
\label{fig:angles}
\end{figure}

The geometry of the input beams and emitted photons for the traps is shown in the schematic, Fig. \ref{fig:angles}.
The excitation and cooling beams at 493 nm are delivered colinearly at an angle of $\beta_q=45 ^{\circ}$ degrees to the axial ($z$) axis of both traps. The emission direction is perpendicular to the $z$-axis and at an angle $\alpha_q$ to each trap's principal $x$-axis.

We are interested in the angles $\theta_{qi}$ between the excitation/cooling wavevectors and the principal axes $i$ as well as the angles $\psi_{qi}$ between the emission wavevector and the principal axes:

\begin{align}
\cos \theta_{qx} &= -\sin{\beta_q} \sin{\alpha_q}   &\cos\psi_{qx} &= \cos\alpha_q\\
\cos \theta_{qy} &= \sin{\beta_q} \cos{\alpha_q}   &\cos\psi_{qy} &= \sin\alpha_q\\
\cos\theta_{qz} &= -\cos\beta_q                   &\cos\psi_{qz} &= 0.
\end{align}

Hence, the Lamb-Dicke recoil parameters are given by
\begin{align}
 \eta_{qi}&=\sqrt{\frac{\hbar k^2}{2m\omega_{qi}}}|\cos\psi_{qi}-\cos\theta_{qi}| \\
\zeta_{qi}&=\sqrt{\frac{\hbar k^2}{2m\omega_{qi}}}|\cos\psi_{qi}|   
\end{align}
and the expected Doppler cooling limit is\cite{Wineland1979} 
\begin{align}
\bar{n}^{\mathrm{D}}_{qi} = \frac{\gamma}{4\omega_{qi}}\left[\frac{\Delta}{\gamma}+\frac{\gamma(1+s)}{4\Delta} \right]\left(1+\frac{1}{3\cos^2 \theta_{qi}}\right).
\end{align}
Here $\Delta \sim \gamma/2$ is the red detuning of the Doppler cooling beam from resonance and $s=I/I_{\mathrm{sat}}\sim 2$ is the saturation parameter with $I$ the laser intensity and $I_{\mathrm{sat}}$ the saturation intensity.
We note that one particular principal axis in Bob (B) is nearly orthogonal to the cooling beam ($\theta_{By}\approx 87^\circ$). This single direction has a poor Doppler cooling limit and hence is the dominant source of error in Eq. \ref{eqn:FidLim}. We estimate an uncertainty in all geometrical angles to be $< 3^{\circ}$.

\begin{table}[ht!]
\caption{Summary of beam angles, Lamb-Dicke recoil parameters, and Doppler cooling limits $\bar{n}^{\mathrm{D}}_{qi}$ for each motional normal mode $i$ of ion $q$.}
\begin{tabular}{ |c|c|r|r|r|r|r|r|r|r| }
 \hline
  $q$& $i$& $\frac{\omega_{qi}}{2\pi}$ (kHz) & $\alpha_q$\phantom{0}& $\beta_q$\phantom{0}& $\theta_{qi}$\phantom{0}& $\psi_{qi}$\phantom{0}& $\eta_{qi}$\phantom{0} & $\zeta_{qi}$\phantom{0}& $\bar{n}^{\mathrm{D}}_{qi}$ \\
  \hline
    \hline
 A & $z$ &  991.5& -\phantom{0}  & 45$^\circ$ & 135$^\circ$\phantom{.8}& 90$^\circ$\phantom{.5}& 0.055 & 0\phantom{00}     & 13\\ \hline
 A & $x$ & 1157.5& 45$^\circ$\phantom{.5}& 45$^\circ$ & 120$^\circ$\phantom{.8}& 45$^\circ$\phantom{.5}& 0.086 & 0.051\phantom{7} & 15\\ \hline
 A & $y$ & 1488.0& 45$^\circ$\phantom{.5}& 45$^\circ$ & 60$^\circ$\phantom{.8}& 45$^\circ$\phantom{.5}& 0.013 & 0.045\phantom{7} & 12\\ \hline
   \hline
 B & $z$ & 330.3 & -\phantom{0} & 45$^\circ$ & 135$^\circ$\phantom{.8}& 90$^\circ$\phantom{.5}& 0.095 & 0\phantom{00}    & 38\\ \hline
 B & $x$ & 826.7 & 85.5$^\circ$ & 45$^\circ$ & 134.8$^\circ$ & 85.5$^\circ$ & 0.066 & 0.0067 &15\\ \hline
 B & $y$ & 992.0 & 85.5$^\circ$ & 45$^\circ$ & 86.8$^\circ$ & 4.5$^\circ$  & 0.073 & 0.077\phantom{7}  & 826\\ \hline
\end{tabular}
\label{table:modes_supplementary}
\end{table}

\subsection{Error budget}
\label{sec:error}
The largest source of error in the entangled state fidelity is intensity fluctuations in the 1762 nm laser that drives coherent qubit rotations, contributing to SPAM and the swap of qubit states in the protocol.
This and several additional sources of errors and noise are summarized in Table \ref{table:errorbudget}.
The temporal wavefunctions of the photons are matched to within 30 ps at the BS by equalising the path length between the excitation laser and the BS, leading to a small error of $0.004$.
We expect a fidelity error of $0.2\%$ from residual entanglement with ion motion due to the recoil from the spread of photon detection times within each time-bin, at a detection window of $\delta t = 10$~ns (see Fig. \ref{fig:deltat-fidelity}).
Dark counts on the photon detectors and background scattered light from the excitation pulse are expected to contribute $<$0.2\%. 
Imbalance in the fiber BS (measured to be less than 2$\%$ from the nominal 50:50) and imperfect BS mode matching are expected to limit fidelity errors to below $0.1\%$. 
We observe a differential qubit coherence time of $2.1$ ms, likely due to  differential magnetic field noise between the two qubits.
This is expected to reduce the fidelity by $< 10^{-4}$ during the $\sim8$ $\mu$s dwell time between the early photon detection and the analysis $\pi/2$-pulse (see Supp. Info. \ref{sec:T2}).
Residual rf micromotion of trapped ions \cite{Berkeland1998} can result in a fluctuating frequency of the emitted photons, causing a phase error in the final entangled state and a reduction in the fidelity.  We measure a micromotion-induced Doppler shift of under 200 kHz through a photon autocorrelation procedure \cite{Berkeland1998}, and expect this to contribute to a fidelity error of less than $10^{-4}$. 

\begin{table} [h]
    \centering
\caption{Sources of error affecting fidelity and their magnitudes.}
\label{table:errorbudget}
    \begin{tabular}{|l|r|} \hline 
          & \textbf{Fidelity}\\ 
         \textbf{Source of error} & \textbf{Error\hspace{2mm} }\\ \hline \hline
         SPAM / 1762 intensity fluctuations     & 0.01\phantom{00}\\ \hline 
         Photon wavepacket overlap              & 0.004\phantom{0} \\\hline 
         Atom recoil, $\delta t = 10$~ns        & 0.002\phantom{0} \\ \hline
         Background counts                      & $<0.002$\phantom{0}\\ \hline 
         Atom recoil, $\omega_{qi}$ fluctuation & $<0.001$\phantom{0} \\ \hline
         Beamsplitter imperfection              & $<0.001$\phantom{0}\\ \hline
         Residual erasure errors                & $<0.001$\phantom{0} \\ \hline
         Micromotion                            & $<0.0001$\\ \hline 
         Coherence time                         & $<0.0001$\\ \hline \hline        
         \textbf{TOTAL}                         & \textbf{0.02}\phantom{00} \\ \hline
    \end{tabular}

\end{table}

\end{document}